\documentclass[11pt]{article}
\usepackage{psfig}
\title{Can spintronic field effect devices compete with their electronic 
counterparts?}
\author{S. Bandyopadhyay\\
Department of Electrical Engineering and Department of Physics \\ 
Virginia Commonwealth University, Richmond, Virginia 23284, USA\\
\\M. Cahay\\
Department of Electrical and Computer Engineering and Computer Science \\ 
University of Cincinnati, Cincinnati, Ohio 45221, USA}

\begin{document}

\maketitle

\begin{abstract}
Current interest in spintronics is largely motivated by a belief
that spin based devices (e.g. spin field effect transistors) will be faster and 
consume less 
power than their electronic counterparts. Here we show that this is generally 
untrue. Unless materials with extremely strong spin orbit interaction can be 
developed, the spintronic devices will not measure up to their electronic 
cousins. We also show that some recently proposed modifications of the original 
spin field effect transistor concept of Datta and Das [Appl. Phys. Lett., 
\underline{56}, 665 (1990)] actually lead to worse performance than the original 
construct.
\end{abstract}

\pagebreak

A spate of device proposals have appeared over the last decade articulating spin 
based analogs of conventional field effect or bipolar junction 
transistors. The field effect variety is motivated  by a seminal concept 
due to Datta 
and Das \cite{datta} who proposed an electronic analog of the electro-optic 
modulator. The Datta-Das device consists  of a quasi 
one-dimensional 
semiconductor 
channel with ferromagnetic source and drain contacts (Fig. 1). Electrons are 
injected with a definite spin orientation from the source,  
which is then controllably precessed in the channel with a gate-controlled 
Rashba spin-orbit
interaction \cite{rashba}, and finally sensed 
at the drain. At the drain end, the electron's transmission probability depends 
on the relative 
alignment of its spin with  the drain's (fixed) magnetization.  By controlling 
the angle of spin precession in the channel with a gate voltage, 
one can control the relative spin alignment at the drain end, and hence control 
the 
source-to-drain  current. This realizes the basic ``transistor'' action. Because 
of this attribute, the Datta-Das device came to be known as the Spin Field 
Effect Transistor (SPINFET) even though its original inventors aptly termed it 
an analog of the electro-optic modulator (not a ``transistor'').

\begin{figure}
\centerline{\psfig{figure=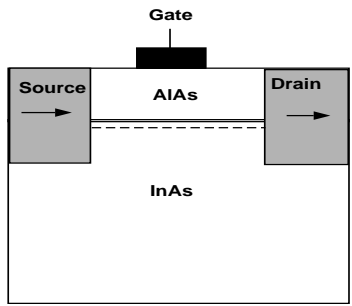,height=5in,width=5in}}
\caption{Schematic of a Spin Field Effect Transistor (or more aptly a spintronic 
analog of an electro-optic modulator)}
\end{figure}

There are many incarnations of the SPINFET
(see, for example, \cite{loss1,loss2,flatte}). All of them however rely on the 
basic concept of modulating the transistor's source to drain current by varying 
the Rashba interaction in the channel with a gate voltage. Therefore, the 
present analysis is perfectly general and applies to all of them. We show that 
in terms of common performance metrics (power dissipation, transconductance, 
unity gain frequency, etc.), the performance projections for a SPINFET are 
below those for a conventional silicon or GaAs field effect transistor.

The following analysis applies to a SPINFET with a strictly one-dimensional 
(1-d) channel. The 1-d SPINFET is the ideal device with the best possible 
performance for two very important reasons. The first reason was 
identified in \cite{datta} itself; one dimensional carrier confinement  
eliminates the angular spread in the electron's wavevector, which results in the 
strongest conductance modulation. In fact, only in a strictly 1-d channel,  the
``off'' conductance of the device can fall to zero resulting in no leakage 
current in the off state. This is extremely important to avoid standby power 
dissipation if two SPINFETs, one biased in the positive transconductance region 
and another in the negative transconductance region, are connected in series to 
act like a complementary metal oxide semiconductor field effect transistor 
(CMOS). The present dominance of CMOS in virtually all electronic circuits is 
due the property that there is no standby power dissipation because the 
leakage current in a conventional MOS transistor is virtually zero when it is 
turned off.
Therefore, at the very outset, it is obvious that only a 1-d SPINFET can have 
any chance of competing with present day silicon CMOS devices. The second 
reason to prefer a strictly 1-d channel is that the major spin relaxation 
mechanism in the channel (D'yakonov-Perel') can be completely eliminated if 
transport is single channeled
\cite{cond-mat}. Therefore, a 1-d channel is always optimum.

The maximum conductance of a strictly 1-d channel is 2$e^2/h$. Since the drain 
current in a ballistic 1-d channel
will saturate when the source-to-drain bias $V_{SD}$ becomes equal to $E_F/e$ 
($E_F$ is the 
Fermi energy in the channel), we have 
\begin{equation}
I_D|_{sat} = 2e E_F/h
\end{equation}
The switching voltage $V_{s}$ to turn the SPINFET from the ``on'' state to 
the ``off'' state is the gate voltage required to precess the spin in the 
channel through an angle of $\pi$ radians. Using the result of ref. 
\cite{datta}, this voltage is 
\begin{equation}
V_{s}|_{SPINFET} \approx \hbar^2 \pi/(2 m^* L \zeta )
\end{equation}
where $m^*$ is the effective mass of the carrier in the channel, $L$ is the 
channel length, and $\zeta$ is a proportionality constant that describes the 
gate 
voltage dependence of the Rashba coupling constant $\eta$. We can theoretically 
estimate $\zeta$. According to ref. \cite{pikus,dietl}
\begin{equation}
\eta = {{\hbar^2}\over{2m^*}} {{\Delta (2 E_g + \Delta)}\over{E_g(E_g + \Delta) 
(3E_g + 2 \Delta)}} {{2 \pi e^2 N_s}\over{\kappa}}
\label{eta}
\end{equation}
where $e$ is the electronic charge, $E_g$ is the bandgap, $\Delta$ is the spin 
orbit splitting in the valence band, $\kappa$ is the static dielectric constant  
and $N_s$ is the surface electron concentration at the interface of the channel 
($N_s$ is related to 
 the interfacial electric field in the channel inducing a structural inversion 
asymmetry and the Rashba effect). From standard MOS theory, $eN_s$ = 
$(\kappa/d)(V_G - V_T)$ where $d$ is the thickness of the gate insulator, $V_G$ 
is the gate voltage and $V_T$ is the threshold voltage to induce an inversion 
layer charge in the channel. Using this result in Equation (\ref{eta}), we find 
that 
 \begin{equation}
 \zeta = {{\partial \eta}\over{\partial V_G}} = {{\hbar^2}\over{2m^*}} {{\Delta 
(2 E_g + \Delta)}\over{E_g(E_g + \Delta) (3E_g + 2 \Delta)}} {{2 \pi e}\over{d}}
 \label{zeta}
 \end{equation}
We will assume an InAs channel and use material parameters from ref. \cite{vur}.
To compare with experiment \cite{nitta}, we will assume that $d$ = 20 nm.
This yields the theoretical value of $\zeta$ = 5$\times$10$^{-29}$ C-m. Equation 
(\ref{zeta}) predicts a linear dependence of $\eta$ on the gate voltage $V_G$. 
Experimentally, one finds the same {\it linear} dependence \cite{nitta}, and the 
experimentally observed value of $\zeta$ $\approx$ 8$\times$10$^{-31}$ C-m 
\cite{nitta}. The theoretical value is about 60 times larger than the 
experimental value, indicating that further experiments are required.

We will now compare the switching voltage of a 1-d SPINFET with that of a 
traditional 1-d MOSFET.
At low temperatures, the switching voltage of a traditional ideal MOSFET (the 
voltage 
required to deplete the channel of all carriers) is 
$E_F/e$. Therefore,
\begin{equation}
{{V_{s}|_{SPINFET}}\over{V_{s}|_{MOSFET}}} \approx {{\hbar^2 \pi e}\over{(2 
m^* 
L \zeta E_F)}}
\end{equation}
 
In order to maintain single subband occupation, we will assume that $E_F$ is 
less than the energy separation between subbands, which is about 
3 meV in InAs 1-d channels \cite{dietl}. Then,  the SPINFET 
will have a lower switching voltage than a traditional FET {\it only if} its
 channel length $L$ $>$ 4.88 $\mu$m. In calculating this, we assumed the 
theoretical value of $\zeta$. If we had assumed the experimental value instead, 
$L$ has to be larger than 293 $\mu$m!. Therefore, it is obvious that for any 
sub-micron channel length (let alone nanoscale devices), the 
SPINFET will have a much higher switching voltage than a traditional MOSFET. 
This 
immediately shows that the SPINFET is {\it not} a {\it lower} power device, 
contrary to 
popular belief (the dynamic power dissipated during switching a transistor is 
proportional to 
the square of the switching voltage). 

It is of course obvious that we can decrease the switching voltage of a SPINFET 
by decreasing the gate insulator thickness $d$. In Si/SiO$_2$ technology, gate 
insulator thicknesses approaching 1 nm is possible without causing significant 
gate leakage, but that may not be possible in systems such as AlAs/InAs (where 
the lower gap semiconductor is chosen for strong Rashba coupling) because the 
barrier height between the semiconductor and insulator is not nearly as high.
We may be limited to a gate insulator thickness of 5 nm or larger in the 
AlAs/InAs system, which still
 makes the switching voltage of a sub-micron SPINFET larger than that of a 
sub-micron MOSFET. Reducing the gate insulator thickness also has deleterious 
effects on the unity gain frequency since it increases the gate capacitance (see 
Equation (\ref{ft}) later).
 
Next, we consider the transconductance of a SPINFET. This is an important 
parameter since it determines device amplification, as well as bandwith or, 
equivalently, device speed. The transconductance of the SPINFET is
 \begin{equation}
 g_m \approx I_D|_{sat}/V_{s} = 2e E_F m^* L \zeta/(\pi^2 \hbar^3)
 \end{equation}
where we have assumed that $V_{s}$ is small enough that $E_F$ does not
vary significantly as the gate voltage swings over an amplitude of $V_{s}$. 
The above equation yields $g_m$ = 6.5$\times$10$^{-6}L$ Siemens (where $L$ is 
the channel length expressed in microns). It is actually more meaningful to 
calculate the transconductance per unit channel width since in conventional 
MOSFETs, the transconductance is proportional to 
the channel width.  For a 1-d channel, we will assume that the confinement 
potential along the width is parabolic, so that the effective width of the 
channel is given by $W_{eff}$ = $\sqrt{\hbar/( 2 m^* 
\omega)}$ \cite{merzbacher}. Since $\hbar \omega$ = 3 meV, $W_{eff}$ = 22 nm.
Therefore, the transconductance per unit channel width is 295$L$ mS/mm, where, 
once again, $L$ is expressed in microns. For sub-micron channel lengths, $g_m$ 
$<$ 295 mS/mm, which is considerably less than what is achieved with GaAs 
high electron mobility transistors.

The unity gain frequency $f_T  \leq g_m/(2 \pi C_g)$, where $C_g$ is the gate 
capacitance given by $C_g$ = $\kappa_i \epsilon_0 L W_{eff} /d$ ($\kappa_i$ is 
the relative dielectric constant of the gate insulator). Accordingly,
\begin{equation}
f_T  \leq 2e E_F m^* d \zeta/(2 \pi^3 \kappa_i \epsilon_0 \hbar^3 
W_{eff})
\label{ft}
\end{equation}

 We will 
assume that the gate insulator is AlAs (relative dielectric constant $\kappa_i$ 
$\approx$ 8.9 \cite{lockwood}) and that $d$ = 20 nm, as before. Using these 
values in Equation (\ref{ft}), we find that $f_T$ $\leq$ 30 GHz. This is less 
than what has 
already been demonstrated for GaAs MESFETs \cite{sia}.

We will conclude this Letter by examining two recently proposed modified 
versions of the SPINFET that claimed to  provide better performance than the 
original proposal of ref. \cite{datta}. The first version \cite{loss1} purports 
to replace a strictly 1-d channel,  where only the lowest subband is occupied, 
with a {\it quasi} 1-d channel where {\it two}
subbands are occupied, in order to provide better spin control. We find this to 
be completely counter-productive for many reasons. First, multi-channeled 
transport (where two subbands are occupied) will not eliminate D'yakonov-Perel' 
spin relaxation; that can happen only in strictly single channeled transport 
\cite{cond-mat}. Therefore, a two-subband device is more vulnerable to spin flip 
scattering, which results in degraded device performance. Second, the presence 
of two occupied subbands can result in 
spin-mixing effects \cite{governale} that are harmful for the SPINFET. Third, 
multiple gates are required in the proposal of ref. \cite{loss1} for conductance 
modulation, and these gates have to be synchronized precisely in order to turn 
the device off. This is an additional engineering challenge that was not 
required in the original proposal of ref. \cite{datta}.

Another type of SPINFET that claims to be able to release the requirement of 
ballistic transport, which is necessary in the original Datta-Das device, has 
recently been 
proposed \cite{loss2}. The idea here is to balance the Rashba interaction 
\cite{rashba} with the Dresselhaus interaction \cite{dresselhaus} (using a gate 
to tune the Rashba interaction). When they are exactly balanced, the 
eigenspinors in the channel are $[1, \pm exp(i \pi/4)]$ which are spins 
polarized on the x-y plane subtending an angle of $\pi/4$ with the x- or y-axis. 
In the convention of Miller indices, we call this axis the [1 1 0] axis. Then, 
by using a ferromagnetic source
contact that is magnetized in the [1 1 0] direction, one can inject all spins 
into one of the eigenstates. Such a spin will traverse the channel without 
flipping (unless there are magnetic scatterers) since it is an eigenstate in the 
channel. However when the gate voltage is detuned to unbalance the Rashba and 
Dresselhaus interactions, the eigenspinors are no longer $[1, \pm exp(i 
\pi/4)]$, but become wavevector dependent. Therefore, any non-magnetic scatterer 
(impurity, phonon, etc.) which changes the electron's wavevector, can also flip 
the 
spin. A spin injected in the [1 1 0] direction is no longer an eigenstate 
and will flip in the channel. The drain is also magnetized in the [1 1 0] 
direction, which will not transmit the flipped spin. Therefore, the device 
conductance will decrease. This device is ``on'' when the gate voltage exactly 
balances the Rashba and Dresselhaus interactions, and `off'' otherwise.

It is difficult to calculate the off conductance of this device since that 
depends on the frequency and nature of spin flip scatterings that occur when the 
Rashba and Dresselhaus interactions are unbalanced. However, it is obvious that 
the off-conductance {\it  is not zero}. In fact, if the device is long enough, 
then a spin arriving at the drain contact is equally likely to be parallel or 
anti-parallel to the drain's magnetization. Therefore, the minimum value of the 
off-conductance in a long-channel device is one-half of the on-conductance. In a 
short-channel device, the minimum value of the off-conductance is even larger. 
Such a device 
is not suitable as a transistor in digital applications (since the on- and 
off-states are not well separated) and even for analog applications, the device 
is less preferable to the original Datta Das proposal since the transconductance 
of this device will be roughly one-half of the transconductance of the Datta-Das 
device. Most importantly, this device has a large leakage current during the 
off-state (at least one-half of the on-current). Therefore, such devices 
will lead to unacceptable standby power dissipation. 

Recently, we have proposed a different type of spin field effect transistor
based solely on the Dresselhaus interaction \cite{cond-mat2}. While it may have
some slight advantages over other renditions of spin field effect transistors, 
it is also not likely to be superior to an ideal 1-d MOSFET in terms of speed
or power dissipation.

In conclusion, we have shown that {\it present versions} of spin based field 
effect transistors are not likely to be competitive with their electronic 
counterparts in terms of speed or power dissipation.  We have also shown that 
some recently proposed improvements over the original 
Datta-Das device of ref. \cite{datta} are actually counter-productive. It is 
therefore unlikely that present versions of spintronic field effect transistors  
will play a 
significant role in combinational digital, analog or mixed signal circuits. 
However, they certainly can play a role in memory (where high gain, high 
frequency, etc. are not necessary). Spintronic devices may also have better 
noise performance since spin does not easily couple to stray electric fields 
(unless the host material has very strong spin orbit interaction). It is also 
possible that
spintronics can outpace electronics in non-conventional applications 
such as single spin logic \cite{bandy,molotkov,bychkov}, spin neurons \cite{wu} 
and using spin in a quantum dot to encode qubits 
\cite{bandy1,loss3,bandy2,khitun}.

\paragraph{Note added:} After the submission of this paper for publication, we 
became aware of a paper by M. Dyakonov that questions the promise of spintronics 
(www.arXiv.org/cond-mat/0401369). Our conclusions in this paper however are only 
specific to spin field effect transistor.

\pagebreak

\end{document}